\documentclass[reprint, floatfix]{revtex4-1}

\usepackage{graphicx}
\usepackage{amssymb}
\usepackage{amsmath}
\usepackage{epsfig}
\usepackage{chemarr}
\usepackage{url}
\usepackage{natbib}

\usepackage{dcolumn}% Align table columns on decimal point
\usepackage{bm}% bold math
\usepackage{color}

\usepackage{comment}
\usepackage[]{changes}
\definechangesauthor[name={Arvind's remarks}, color=red]{Arvind}
\definechangesauthor[name={Nachi's remarks}, color=orange]{Nachi}
\definechangesauthor[name={Viraaj's remarks}, color=orange]{Viraaj}

\begin{document}

\title{Shaping dynamical folding and misfolding pathways in mechanical metamaterials}

\author{Menachem Stern, Viraaj Jayaram, Arvind Murugan}

\affiliation{Department of Physics and the James Franck Institute,
	University of Chicago,	929 E 57th Street, Chicago IL 60637}

%\date{\today}

% Main points to make:
% Pathways - a sequence of states.. 
% Dynamics - traversed at a finite rate
% 
\begin{abstract}
The design of desired behaviors in mechanical metamaterials has produced remarkable advances but has generally neglected two aspects - the inevitable presence of undesired behaviors and the role of dynamics in avoiding such behaviors.  Inspired by similar hurdles in molecular self-assembly and protein folding, we derive design principles to shape dynamical folding and misfolding pathways in disordered mechanical systems.  We show that such pathways, i.e., sequences of states traversed at a finite rate, are determined by the bifurcation structure of configuration space which, in turn, can be tuned using imperfections such as stiff joints.
%by programming the bifurcation structure of
%two new ideas; (a) design parameters should aim `off-target' from the desired behavior to better suppress undesired behaviors, and (b) such design principles should target entire dynamic folding and misfolding pathways and not just the final states. 
We apply these ideas to completely eliminate the exponentially many ways of misfolding a self-folding sheet by making some creases stiffer than others.
%In particular, we show that stiff springs, when placed along creases predicted by a linear programming problem, dramatically alleviates misfolding in the very slow or the very fast folding limit as desired. 
Our approach also shows how folding at different rates can controllably target different desired behaviors.
\end{abstract}
\pacs{}
\maketitle

The field of mechanical metamaterials has sought to program the micro-structure of materials with specific properties in mind. This approach has achieved numerous successes, ranging from the targeted design of exotic elastic moduli and non-linear responses to achieving specific geometries \cite{Peraza-Hernandez2014-gp,silverberg2014,Pellegrino:2014vg,Reis:2015ey,bertoldi2017flexible}. But these approaches are also faced with limitations, where certain properties or geometries are not easily targeted through currently available design principles \cite{bertoldi2017flexible,Dudte2016-ws,Paper2,chen2018branches}.

In this work we explore a new design principle inspired by similar problems in biology. We argue that design parameters should not be chosen based on just the desired behavior but also to minimize undesired behaviors, even at the expense of distorting the desired behavior. Further, desired and undesired behaviors should not be viewed as isolated states but as the end point of \textit{dynamical pathways} -- an entire sequence of states that are traversed at a finite rate. 

While such ideas are common place in protein folding and self-assembly of macromolecular structures and viruses \cite{Deeds2007-bl,Murugan2015-wb,Jacobs2017-jv,Dobson2003-og}, the role of undesired structures and that of dynamical pathways have not been systematically explored in meta-materials design.

%Further, any actuation process for a complex metamaterial, that might superficially look like an instantaneous switch between two states, involves a \textit{dynamical folding pathway} in practice -
%Undesired states must be lifted by distorting the desired one..

%Taken together, we argue that design parameters should target intermediates that favor desired structures more than undesired ones; such intermediates may not resemble the desired structure and yet are optimal at promoting the ultimate odds of success. 

% Need to say what the dynamical pathways are..  given by the bifurcation structure. 

 We introduce these ideas here in materials with multiple low (or zero) energy modes meeting at an unstrained state. We find that introducing  stiffness in parts of the material (e.g., joints) can change the topological connectivity of these non-linear modes. With the right distribution of stiffnesses, all undesired modes can be arranged to disappear in saddle-node bifurcations at higher strain and thus not be accessible from the unstrained state. Thus, joint stiffness dramatically changes the topology of mode bifurcations in configuration space.
 %intrinsic energy landscape as a function of the total strain. 
 
The bifurcation diagram, in turn, determines the outcome based on the actuation rate. For example, adiabatic folding results in the structure continuously connected to the unique mode available at low-strain in the bifurcation diagram while fast folding selects the structure most similar to the unique low-strain mode.

%we find that introducing joint stiffnesses will raise the energies of different modes to different extents and can completely eliminate all but one of the modes at low strain. These stiffness create an continuous deformation of the intrinsic energy landscape as a function of the total strain. If the stiffnesses are tuned right, all distractor minima can be arranged to disappear in saddle-node bifurcations, leaving a simple small-strain landscape with a unique minimum. Such a continuous deformation can be helpful if we 

% we show that mechanisms with stiff joints have an energy landscape that is continuously modified as a function of total strain. At large strains, the stiffnesses play little role. However, at small strains, the stiffness 
% We show that with stiffness, these minima continue to exist at large folding but the energy landscape is continuously modified as a function of the total strain. If tuned right, the distractor minima all disappear in saddle-node bifurcations and the small-strain landscape has a unique minimum. Such a landscape helps with adiabatic  folding, e.g., folding much slower than relaxation timescales of the joints in the system. 
% The bifurcation structure also determines the outcome when folding fast; e.g., at very fast folding rates, the unique minimum of the simplified small-strain landscape selects the actuated mode. 

We apply these ideas to the exponential folding problem faced by a self-folding sheet \cite{Lang:2007,Demaine:2007vk,Tachi2009-kh,Tachi:2010tk,Dudte2016-ws}. These sheets have an exponential number of misfolding pathways that meet at a `branch point' at the flat state \cite{Paper2,chen2018branches,chen2011bifurcation,rocklin2018folding}, making it nearly impossible to pick the desired folding mode. Similar `branch point' problems have been explored for a variety of mechanisms in the past \cite{myszka2012mechanism,Myszka2012-ap, Myszka2014-hx}.

Our results have practical implications. Currently, many approaches \cite{Peraza-Hernandez2014-gp} put in effort into introducing directional asymmetry in materials so that, e.g., creases will fold in one way (say, Mountain) but not the other (Valley). This is seen as the way to pick one folding mode over others. Counter-intuitively, our results suggest that non-directional crease stiffness - a symmetric property that cannot distinguish Mountain from Valley and also inevitable in most materials - can pick between distinct global Mountain-Valley configurations. We show that heuristics for the needed stiffnesses can be formulated as computationally easy Linear (and Quadratic) Programming problems, with associated properties like sparsity of stiff creases. 

Our results have broader implications. Our result exploits non-linearity for a functional purpose - strong competing non-linearities allow us to design heterogeneous folding pathways where there are fewer choices for small displacements near the unstrained state and more choices later. By controlling these non-linearities and where bifurcations occur, the speed of folding can select between different target structures.

%Rate dependence and off-target design are closely related. Rate dependence comes in because there is no way to eliminate the undesired minima without distorting the desired structure in some way. The distorted intermediate state can go back to the full structure. 

%In other systems, a hierarchy of stiffness has been shown to substitute for temporal ordering and temporal control. Our approach based on crease stiffness can be seen as a variant of this idea, suggesting that the approach here can also apply to patterns with more degrees of freedom. Examples of this have been seen outside of origami. 
%While applied to self-folding origami, our approach can also apply to temporally ordered folding systems where stiff creases naturally fold less than others. 

%\section{Captions}

\begin{figure}	
\includegraphics[width=\linewidth]{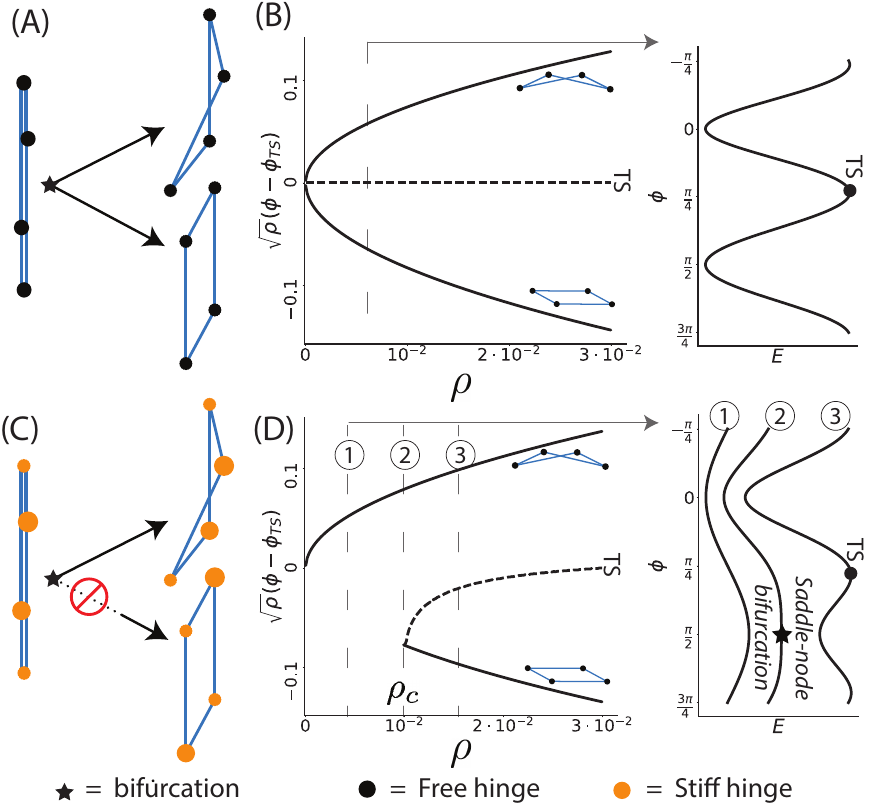}
\caption{Stiff joints in a linkage network can change the connectivity of non-linear modes in state space. (a-b) The 4-bar linkage has only one degree of freedom but two distinct zero energy motions that meet at a branch point at the flat state, making the mechanism difficult to control. The two motions can be seen as minima in the energy landscape at a fixed total strain $\rho$. (c-d) We can eliminate a chosen motion in a saddle-node bifurcation at $\rho_c$ by making the joints stiff to different extents (i.e., adding torsional springs that are relaxed in the flat state). The bifurcation diagram shows that such stiffness have changed the connectivity of the two non-linear modes. One of the two modes is destroyed in a saddle-node bifurcation at $\rho_c$ and is thus inaccessible from the flat state $\rho = 0$. (Here, $\phi$ is the angular variable in the 2-dimensional linearized null space of the flat state.)
%Placing stiff springs on some hinges, resisting rotation away from the flat state, lifts the minimum of the antisymmetric state, making it inaccessible from the original bifurcation point. c) Placing springs on other hinges lifts the symmetric minimum.
\label{fig:Linkage}}
\end{figure}

\begin{figure*}	
\includegraphics[width=\linewidth]{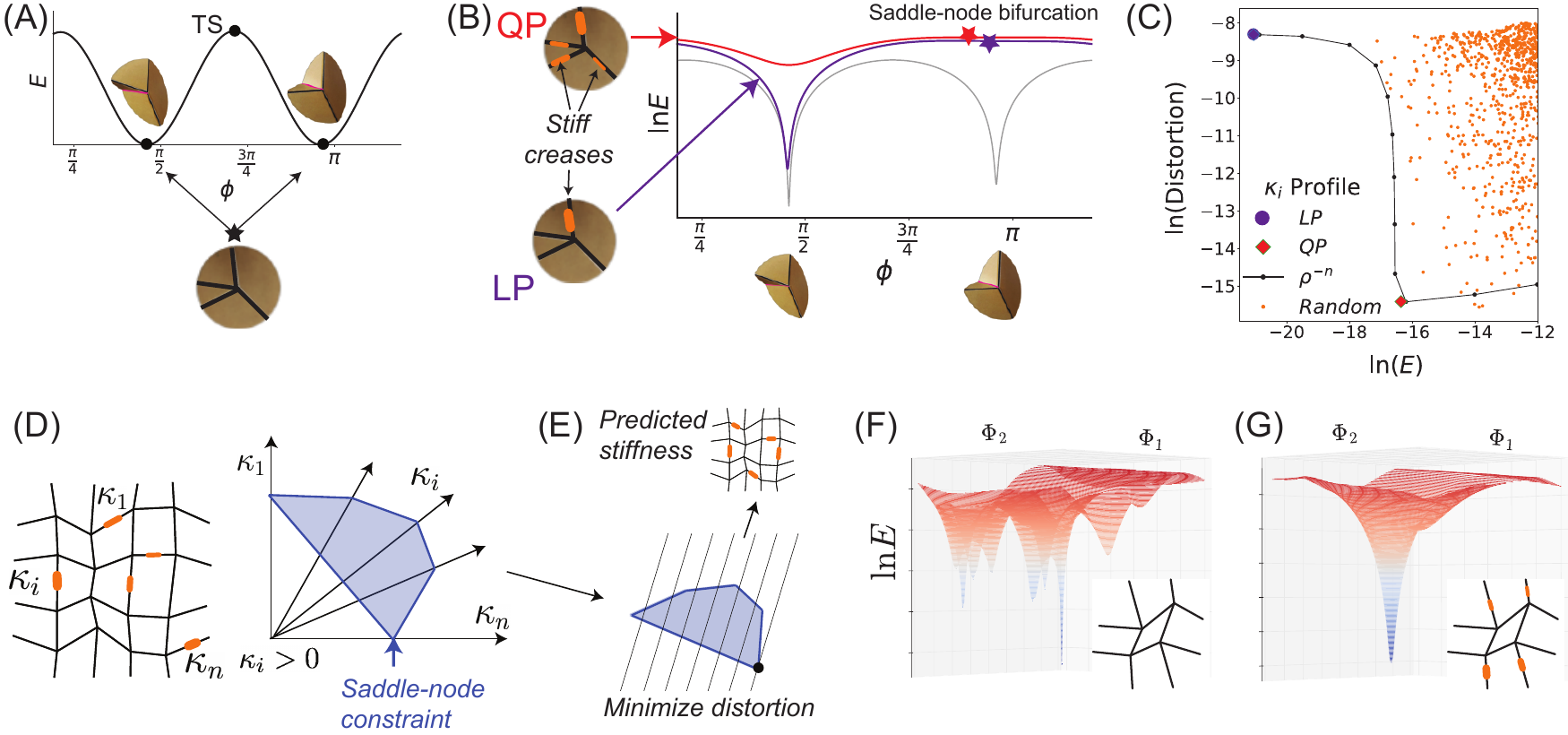}
\caption{Heterogeneous stiff creases can simplify the landscape of self-folding sheets near the flat state. (a) An origami 4-vertex has a choice between two distinct folding modes at the flat state. (b-c) Making creases stiff completely eliminates a chosen mode by combining it with a nearby transition state (TS) in a saddle-node bifurcation. (Thickness of orange strip indicates stiffness.) (c) Trade-off: Stiff creases distort the desired mode while eliminating undesired modes. Stiffness profiles that minimize energy distortion (e.g., Linear Programming (LP) method) cause large geometric distortion and vice-versa (e.g., Quadratic Programming (QP) method). % Crease stiffness prescriptions of $\kappa_i \propto 1/\rho_i^n$, as $n$ is varied, comprise the trade-off front.  
(d) The exponentially many misfolding modes of large sheets are all eliminated if the stiffness profile $\kappa_i$ satisfies a linear constraint, shown here as a simplex. (e) We can minimize distortion (energy or geometry) of the desired mode by optimizing spring stiffness on this simplex. %Linear or quadratic  programming problems are easily solved on a computer; the solution to the former is necessarily a sparse set of crease stiffness. 
(f-g) All but one chosen minimum in a quad pattern's energy landscape (at small overall folding) can be eliminated by stiff creases predicted by procedure in (e). 
\label{fig:VertexQuad}}
\end{figure*}

%The extreme versions are given by a linear programming and a quadratic programming problem respectively. Thus these extreme limits are guaranteed to be sparse; only one crease needs to be stiff to achieve these results.    

\begin{figure*}	
\includegraphics[width=\linewidth]{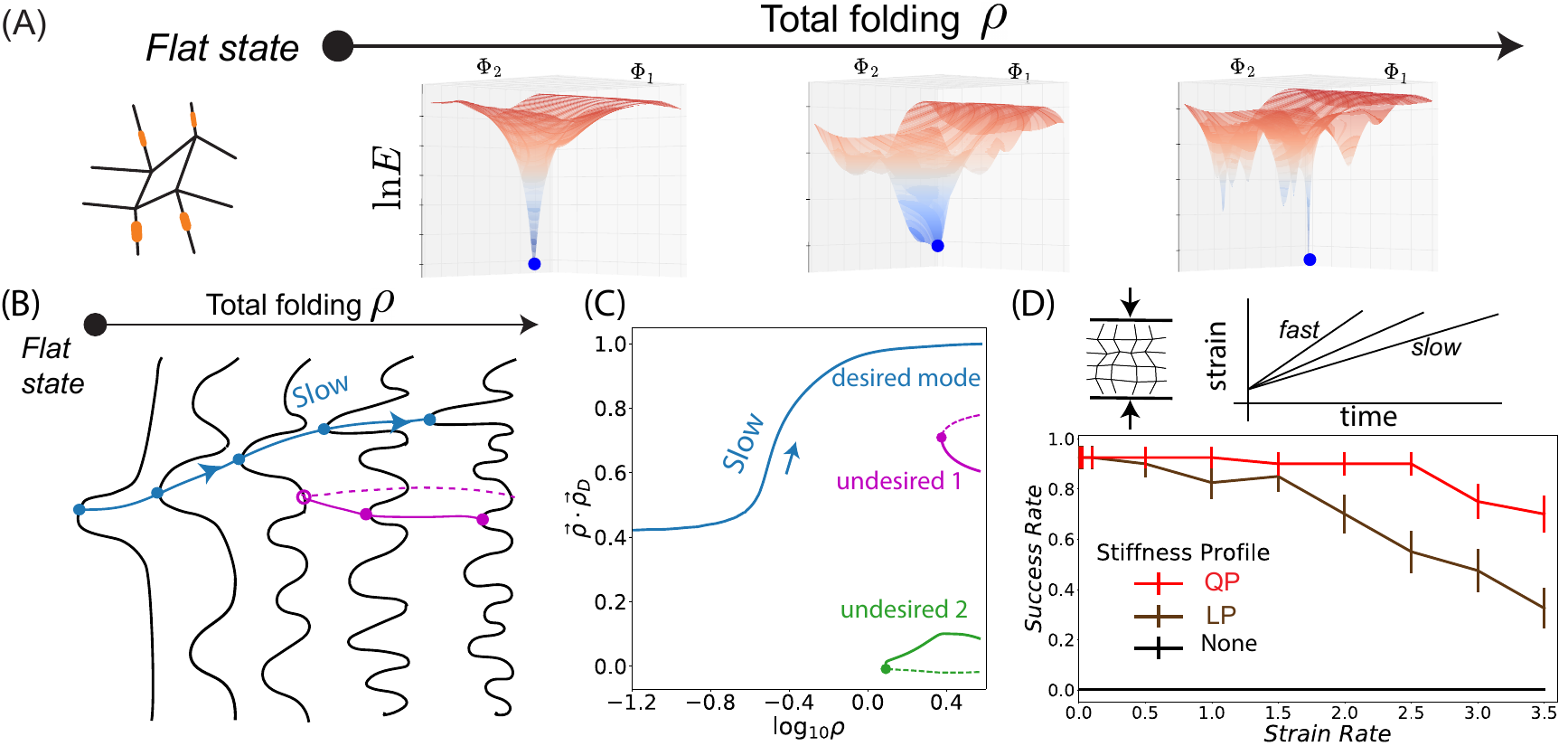}
\caption{Stiff creases change the topological connectivity of undesired modes. (a) The energy landscape of a quad with stiff creases has a unique mode at low strain $\rho$ but becomes more complex at higher strain. (b)  However, slow folding will recover the desired state provided the unique mode at low strain is continuously connected to the desired state and does not undergo a saddle-node bifurcation (blue line). Slow folding can be successful even if the unique mode at small strain $\rho$ is quite distorted relative to the desired mode.  (c) Bifurcation diagram as a function of total folding $\rho$ for a specific $16$-vertex pattern shows select undesired modes (solid lines) being eliminated at bifurcations with saddles (dashed lines). Only the desired mode (blue) survives - albeit distorted - to the flat state. (d) Simulations of folding at a finite strain rate (relative to relaxation timescale of hinges) show high success for slow folding and failures at higher folding rates. %Quadratic Programming (QP)-based heuristics perform better than Linear Programming (LP) at faster folding rates because QP causes less geometric distortion. 
(Data from 50 random $16-$vertex patterns).
\label{fig:bifurcationdiagram}}
\end{figure*}

%Figure 5: (a) Adiabatic folding. Success rate as a function of dot product between folding torque used and desired mode. Results shown are an average over $10$ instances of square patterns made of $5\times 5$ vertices.    
%(b) Fast folding.  Success rate as a function of dot product between folding torque used and desired mode. Results shown are an average over $10$ instances of square patterns made of $5\times 5$ vertices.

\section{Results}

\subsection{Avoided bifurcation in linkage networks}

We first demonstrate our ideas on a simple but canonical model, namely the 4-bar mechanical linkage \cite{hartenberg1964kinematic,mccarthy2006geometric} in Figure~\ref{fig:Linkage}(a,b). While the structure has only one Maxwell degree of freedom, the flat state is a special point - it sits at a bifurcation where the degree of freedom is branched (and associated with a self-stress mode) \cite{chen2014nonlinear}. When compressed as shown, the linkage must choose one of the two distinct zero energy motions that conserve rod lengths. The associated energy landscape, at some fixed compression, has two minima corresponding these motions with an energy barrier (and transition state $TS$) between them; see Figure~\ref{fig:Linkage}(b). (See SI for precise energy model.) Many studies \cite{wampler1986manipulator,Wampler2011-cf,kieffer1994differential} have sought to predict and eliminate such `branch points' in complex mechanisms because one of the modes is usually desired and functional while the other is undesired. 
%We take a different approach and observe that the mechanical advantage (i.e., folding angles at joints) in the distinct motions are different. %We take a different approach and observe that joint stiffness would affect the different modes to different extents. 

We take a different approach and observe that experimental realizations of such mechanisms \cite{howell2001compliant,kota1995designing,Paper1} have imperfections that lift the energies of all the modes shown. If an imperfection can raise energy of the undesired zero mode more than it raises the energy of the transition state $TS$ and that of the desired mode, the undesired mode would disappear in a saddle-node bifurcation with the state at the barrier.

One such imperfection is stiffness in the joints. We model the stiffness of joint $i$ by a torsional spring of stiffness $\kappa_i$ that is relaxed in the flat configuration shown, i.e., at the branch point. That is, we assume a joint energy $E_i =  \kappa_i \rho_i^2/2$, $\rho_i$ being the folding angle measured from the flat state configuration. 

We find that if the joints have unequal stiffness $\kappa_i$, it generically lifts the energies of different modes to different extents. In fact, one of the modes undergoes a saddle-node bifurcation with the transition state $TS$ separating the two modes (see Figure~\ref{fig:Linkage}(c,d)) at a finite folding extent $\rho =\rho_c$ where $\rho \equiv \vert \vert \vec{\rho} \vert \vert$. This distance $\rho_c$ is given by a competition between rod bending (or rod compression in alternative models) at the transition state $\sim K \rho^4$ and the spring energy $\sim \kappa \rho^2$; as shown in the SI, $\rho_c \sim \sqrt{\kappa/K}$. Other choices of $\kappa_i$ can eliminate the other mode.
%\begin{align}
%\rho_c \sim \sqrt{\kappa/K} .
%\label{eq:rc}
%\end{align}

% State main result:
Thus joint stiffnesses change the topological connectivity of undesired modes in state space; see Figure~\ref{fig:Linkage}(d). As a result, the undesired mode can be made inaccessible from the flat state, which now continuously connects with only the desired mode.
If the network is actuated slowly relative to relaxation timescales of the stiff joints, the network will fold into the desired mode and stay in that state even for $\rho > \rho_c$, despite the reappearance of the undesired mode at large $\rho$.

\subsection{Misfolding in self-folding sheets}
Self-folding sheets (or self-folding origami) are structures programmed to have one unique low or zero energy mode \cite{Miura1980-qk,Peraza-Hernandez2014-gp,santangelo2017extreme}. However, self-folding sheets, even when programmed with a single zero energy mode, have been shown to have exponentially many undesirable mis-folding modes accessible from the flat state \cite{Paper2,chen2018branches}. We show how crease stiffness can change the topological connectivity of these modes and leave only the desired folding mode accessible from the flat state.    

\subsection*{Avoided bifurcation in a 4-vertex}
The atomic unit of self-folding origami is a 4-vertex \cite{Huffman1976-rr}. Much like the 4-bar linkage, the 4-vertex has one degree of freedom but the flat unfolded 4-vertex is at a branch point, a meeting point of two distinct folding motions \cite{Waitukaitis2015-rw,Tachi2016-si}, distinguished by the Mountain-Valley states of the creases (Figure~\ref{fig:VertexQuad}(a)) \cite{kawasaki1989,hull1994}.  These two motions are shown as zero energy minima in Figure~\ref{fig:VertexQuad}a using a model of vertex energy presented in the SI, with a transition state $TS$ separating them. This binary choice is the origin of the exponentially many misfolds of large self-folding sheets. 

As with the 4-bar linkage, we wish to lift and eliminate one of the two folding motions, making it inaccessible from the flat state.

%It was previously shown that given a total folding magnitude $\rho\equiv \|\vec{\rho}\|\ll \pi$, an energy landscape is formed such that the two minima (in the angular directions) reside in a smaller subspace, the linear null space of the vertex constraint matrix. 

%Much like with the 4-bar linkage, the 4-vertex has one degree of freedom but 

%This branch point for the 4-vertex leads to an exponential number of undesirable mis-folding modes that meet at the flat state \cite{Paper2,chen2018branches} of large self-folding patterns. 

%Given a pattern, its configuration is completely defined by a folding vector $\vec{\rho}$, assigning dihedral angles to each of the creases. By convention we define the flat state as $\vec{\rho}=0$, with mountain (valley) folds given positive (negative) angle values. 

%The critical observation is that the energy minima (at given $\rho$) scale like $E\sim\rho^4$. We can thus augment the energy model with a $E\sim \rho^2$ term to overwhelm one minimum close enough to the flat state $\rho=0$. Such modification will become unimportant for large $\rho$, and thus the original landscape will be asymptotically recovered. 
We introduce stiffness at the creases, an inevitable feature of most material implementations. We model such stiffness of crease $i$ as a torsional spring with $\rho = 0$ rest angle with energy $E_{Crease\; i} = \kappa_i \rho_i^2/2$. The energy of the origami vertex can be written as,
\begin{align}
E &= E_{Vertex} + E_{Crease} %\nonumber \\
%&= E_{Vertex} + \frac{1}{2}\sum_{{i\in \text{creases}}}\kappa_i \rho_i^2 ,
\label{eq:VertexEnergy}
\end{align}
where $E_{Vertex}$ accounts for the vertex constraints \cite{Huffman1976-rr} and $E_{Crease} = \sum_i \kappa_i \rho_i^2/2$ accounts for crease stiffness. For more details on the energy model see SI. Crucially, the vertex energy arises from face bending, and scales with a high power $\rho^4$ for the two special folding motions.

Let us find the conditions on $\kappa_i$ for lifting and eliminating a chosen branch - the `undesired branch' - of the 4-vertex. We assume the folding angles of the undesired mode and the desired mode are $\tilde{\rho}_U$ and $\tilde{\rho}_D$ respectively and that of the transition state $TS$ separating them is $\tilde{\rho}_{TS}$, all assumed to be defined near the flat state and normalized. See Figure~\ref{fig:VertexQuad}(a).

Let $E_{TS}(\rho)$ be the energy of $TS$ at some chosen total folding $\rho\equiv \vert\vert\vec{\rho}\vert\vert$. Here, we will focus on eliminating the undesired minimum up to a distance $\rho_c$ from the flat state and return to larger folding behaviors later. To lift and eliminate the undesired minimum, we should choose a heterogeneous stiffness profile that raises the undesired mode more than the transition state $TS$. This constraint - requiring a saddle-node bifurcation - can be written as,
\begin{align}
\frac{1}{2} \rho_c^2 \sum_{i \in \text{creases}}\kappa_i \Big[(\tilde{\rho}_U)_i^2-(\tilde{\rho}_{TS})_i^2\Big] \geq E_{TS}
\label{eq:VertexLiftingConstraint}
\end{align}

In addition, all crease stiffnesses must be non-negative:
\begin{align}
\kappa_i \geq 0
\label{eq:PositivityConstraint}
\end{align}
Note that both constraints are linear in the stiffnesses $\kappa_i$.

%\subsection*{Minimizing distortion}

Any set $\kappa_i$ satisfying the above constraints will eliminate the undesired mode in a saddle-node bifurcation at a total folding distance $\rho_c$, making it inaccessible from the flat state. 

Only the desired mode is stable in the neighborhood of the flat state but it can be significantly distorted by the stiff creases. As shown in Figure~\ref{fig:VertexQuad}, with stiff creases, the desired mode is of non-zero energy (`Energy distortion') and can also have distorted folding angles (`Geometric distortion'). We wish to formulate design principles for choosing stiffness profiles $\kappa_i$, consistent with the above constraints, that best facilitate designed folding motions. 

%\subsection{Selecting crease stiffness profiles}

%Selecting any crease stiffness profile $\kappa_i$ satisfying the above constraints will maintain only one of the two vertex motions near the flat state. Still, 

We devise two design strategies: (1) Minimizing energy of the desired mode (Energy optimization), (2) Minimizing geometric distortion of the desired mode (Geometric optimization). We find that different crease profiles generally trade-off energy and geometric distortion.

Energy optimization is simple: the desired mode has non-zero energy $E(\vec{\rho}_D)= \sum\kappa_i (\rho_D)_i^2/2$ because of crease stiffness. As this function is linear in $\kappa_i$, optimization subject to the saddle-node constraints Eqs.~(\ref{eq:VertexLiftingConstraint}-\ref{eq:PositivityConstraint}) is equivalent to a Linear Programming (LP) problem:
\begin{align}
\begin{aligned}
& \underset{\bm{\kappa}}{\text{minimize}}
& & E(\vec{\rho}_D) = \frac{1}{2} (\bm{\rho}_D^2)^T\bm{\kappa} \\
& \text{subject to}
& & \rho_c^2\Big[(\tilde{\bm{\rho}}_U^2 )-(\tilde{\bm{\rho}}_{TS}^2)\Big]^T \bm{\kappa} \geq 2E_{TS}\ ,\\
&&& \kappa_i \geq 0, \; i \in \text{creases}\ .
\end{aligned}
\label{eq:LP}
\end{align}
%\textbf{Why factor of 2 $E_{TS}$ here? AM removed it.. explain in SI if used for numeric reasons.}

Linear Programming problems are efficiently solved in polynomial time; further, the optimal stiffness profile $\kappa_i$ is guaranteed to be sparse. In a 4-vertex, only one crease needs to be stiff.

Geometric distortion is minimized if fold angles in the surviving minimum with stiff crease closely corresponds to the fold angles $\vec{\rho}_D$ of the desired branch without stiff creases. Here, we use the gradient of the energy potential with stiff creases, but evaluated at $\rho_D$, as a proxy for such geometric distortion. As shown in the SI, this proxy, after projecting out radial components of the gradient, is: 
%If the gradient of the stiff crease energy is parallel to $\vec{\rho}$, we know that the point is an angular minimum. 
%This condition can be expressed as $\vec{\rho}\cdot\vec{(\kappa\rho})=\rho\|\vec{\kappa\rho}\|$. By squaring this condition  we define a second function to be optimized

%$\frac{\vec{\rho}\cdot\vec{(\kappa\rho})}{\rho\|\vec{\kappa\rho}\|}=1$

\begin{align}
F_{QP} &= \rho_D^2\sum_{{i\in creases}}\kappa_i^2 (\rho_D)_i^2 \ - \sum_{i,j}\kappa_i \kappa_j (\rho_D)_i^2 (\rho_D)_j^2.
\label{eq:QP}
\end{align}

$F_{QP}$ is a positive semi-definite function that is manifestly quadratic in $\kappa_i$. Optimization of this function - with the linear constraints in Eqs.~(\ref{eq:VertexLiftingConstraint}-\ref{eq:PositivityConstraint}) - is facilitated by efficient Quadratic Programming (QP) algorithms.

In practice, the LP and QP prescriptions do well at optimizing their respective strategies (i.e. energy and geometry) for a single vertex. Figure~\ref{fig:VertexQuad}(c) shows how these prescriptions indeed do better than choosing random stiffness profiles that satisfy the constraints. 

Regardless of which optimization prescription is used, the saddle-node constraint Eqnuation~(\ref{eq:VertexLiftingConstraint}) ensures that close to the flat state, the energy landscape in the angular variables becomes convex (up to $Z_2$ symmetry), permitting just a single minimum. This guarantees that, irrespective of the choice of folding protocol, the resulting configuration at small $\rho$ is predictable and reproducible.
%that folds the pattern to small $\rho$, the resulting configuration is completely predictable and reproducible.

\subsection*{Stiffness profiles for large self-folding sheets}

Large origami patterns have exponentially many distractor minima states, making it near impossible to fold correctly \cite{Paper2,chen2018branches}. Still, the ideas of the previous section can be used to lift all but one of these minima at small folding angles. Crucially, the desired self-folding motion of a large pattern \cite{Tachi:2012,Paper1} is consistent with exactly one of the two folding modes for each of its constituent 4-vertices. Thus, for a pattern with $V$ vertices, the saddle-node constraint in Equation~(\ref{eq:VertexLiftingConstraint})  generalizes to $V$ linear constraints, one for each vertex $v$:
%In patterns with $V$ vertices, this furnishes $V$ dependent linear constraints, 

\begin{align}
%\nonumber
\rho_{c}^2 \sum_{i \in \text{creases of }v}\kappa_i \Big[(\tilde{\rho}_{U,v})_i^2-(\tilde{\rho}_{TS,v})_i^2\Big] \geq 2 E_{TS,v}
\label{eq:PatternLiftingConstraint}.
\end{align}

Note that the constraints are dependent since vertices share creases. These linear constraints, along with $\kappa_i >0$, define a simplex in the space of crease stiffnesses as shown in Figure~\ref{fig:VertexQuad}(d). We can still use LP and QP algorithms as before to find optimized stiffness profiles. 

Figure~\ref{fig:VertexQuad}(f-g) shows how applying a LP stiffness profile to a quad  pattern lifts all but one minimum close to the flat state. Sampling a large amount of large patterns shows that LP and QP indeed optimize their respective strategies; {see SI.}

\begin{figure*}	
\includegraphics[width=\linewidth]{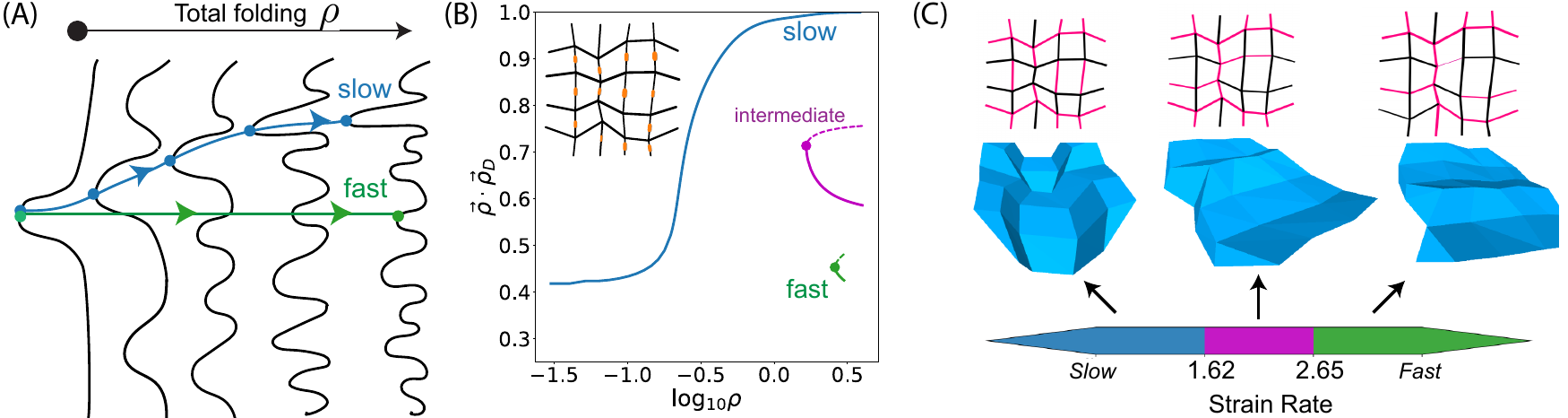}
\caption{Folding rate can controllably select between different folding pathways. (a) While slow folding follows the continuous deformation of the unique mode at low strain $\rho$ (blue), fast folding results in a state that most `resembles' that low-$\rho$ mode (green). If the unique low-$\rho$ mode is  significantly distorted in geometry relative to the slow folding target, slow and fast folding can result in very different outcomes. (b-c) We systematically attempted folding at different strain rates (relative to a fixed hinge relaxation timescale) for the $16$-vertex pattern with stiff creases shown. We find three distinct outcomes at slow, intermediate, fast rates that completely differ in their Mountain-valley states, geometry and energy. The slow folding outcome corresponds to following the blue path in (b) while the intermediate and fast pathways cross over from blue to the green and orange modes respectively at some intermediate folding angles.   
\label{fig:RateDependence}}
\end{figure*}

%\begin{align}
%\Big[(\bm{\rho}_{U,v}^2 )-(\bm{\rho}_{TS,v}^2)\Big]^T \bm{\kappa} \geq 2 %\bm{E}_{TS,v}
%\label{eq:PatternLiftingConstraint2}
%\end{align}

%The LP algorithm continues to supply sparse solutions, such that only one crease per vertex needs to be stiff. 

\subsection*{Larger folding angles and adiabatic folding}

% Terminology
% Mode - non-linear low-freq mode.. what the bifurc diag shows
% Branch - only for branch point. Dont want to use this too much.
% Motion - zero energy mode.
% Structure, Behavior
% Pathway

% Things to define:
% \rho - scalar - as norm of \rho vector.
% \vec{\rho} - folding angles
% \vec{\tilde{\rho}}_D etc - normalized folding angles of states computed near the flat state.

While folding the quad with stiff creases in Figure~\ref{fig:VertexQuad}(g) retrieved the desired structure, we noticed that folding beyond a certain angle gives rise to many new minima (Figure~\ref{fig:bifurcationdiagram}(a)). To understand this, note that the saddle-node bifurcation constraint, Equation~(\ref{eq:VertexLiftingConstraint}), only ensures the absence of undesired modes up to a total folding $\rho_c$ at which $E_{TS}(\rho_c)$ is computed. %Intuitively, crease stiffness energy only dominates up to a folding angle $\rho_c \sim \sqrt{\kappa_c/\kappa_F}$ since $E_{TS} \sim k_F \rho^4$, with $k_F$ the bending modulus for stiff faces (see SI). 
Intuitively, crease stiffness ($\sim \rho^2$) becomes less important than face bending ($\sim \rho^4$) as folding proceeds and the undesired modes are restored in a series of saddle-node bifurcations. See SI for more on the energy model.

At first sight, the reappearance of undesired modes at large $\rho$ might seem disappointing. However, if folding is carried out adiabatically - i.e., slowly relative to hinge relaxation timescales - these modes do not impact folding at all. Adiabatic folding, by definition, will follow the continuous deformation of the unique low-$\rho$ minimum.

% Condition for successful folding
Thus, for successful adiabatic folding, the only condition is that unique low-$\rho$ mode, is continuously deformed to the desired structure at large $\rho$. See Figure~\ref{fig:bifurcationdiagram}(b).

% Not sure what this paragraph is doing here:
Figure~\ref{fig:bifurcationdiagram}c shows the bifurcation diagram for a $16$-vertex pattern with stiff creases predicted by Linear Programming. The unique low-$\rho$ mode is significantly distorted relative to the desired state (i.e., has low dot product). Nevertheless, this mode is continuously deformed to the desired state along the blue path, which was followed in slow folding simulations. Undesired states, on the other hand, are not continuously connected to low-$\rho$ mode.

%the evolution of the normalized dot product of the  with the desired state desired state $\vec{\rho}_D$ as the total folding $\rho$; this blue path is followed by adiabatic folding.  

To test whether our stiff crease prescriptions are able to consistently create adiabatic pathways, we sampled 50 random patterns, each with a programmed low-energy motion using the loop equations of \cite{Tachi:2012,Paper1}. Such patterns are expected to have $\sim 10^3$ higher energy undesired modes, corresponding to motions that jam close to the flat state. Accordingly, we almost never succeed in folding in the desired low-energy mode with generic folding torques. We then augmented the sampled patterns with stiff creases resulting from LP and QP prescriptions and simulated folding at varying rates. (In simulations, we assume the crease hinges follow a first order equation with a relaxation timescale $\tau_{relax}$; this timescale is known to vary for different material implementations \cite{holmes2011bending}.)

Patterns with either LP or QP stiffness profiles achieve a success rate in excess of $90\%$ when folded slowly (Figure~\ref{fig:bifurcationdiagram}(d)), compared with the expected $<0.1\%$ success rate with perfectly soft creases. Thus our heuristics for crease stiffness are useful for slow folding, yet not perfect. 

% Method of failure
The small fraction of failed cases represent patterns where the unique low-$\rho$ mode and the desired high-$\rho$ mode undergo distinct saddle-node bifurcations at intermediate $\rho$ and thus do not connect up. Such bifurcations are mathematically forbidden if these states are the lowest energy states for all $\rho$. Complex optimization methods that account for details of non-linear energy landscape at all intermediate $\rho$ might be able to better protect from such bifurcations. However, we find that simple heuristics, e.g., based on the energy of low and high-$\rho$  states alone, are sufficient to protect the adiabatic pathway from bifurcations for complex patterns. See SI for more analysis of failures.

\subsection{Folding rate-dependent target structures}

%Folding into the designed motion is successful if the achieved configuration at large $\rho$ values has a dot product $\sim 1$ with the designed motion. 

%This is to be expected, as fast folding disregards the adiabatic nature of the morphing from $\vec{\rho}_*$ to $\vec{\rho}_D$. Fast folding essentially takes the pattern to $\vec{\rho}_*$ at high $\rho$ and then relaxes to the nearest angular minimum. If $\vec{\rho}_*$ is well aligned with $\vec{\rho}_D$, fast folding succeeds, explaining the trend that QP stiffness profiles (designed to minimize geometric distortion) have a better success rate at large folding rates.

We have seen that the unique low-$\rho$ minimum funnels adiabatic folding to the desired state in a glassy landscape, even if the unique low-$\rho$ mode is significantly distorted relative to it. However, the success rate drops with folding rate; see Figure~\ref{fig:bifurcationdiagram}(d).

Such a drop in success rate is to be expected since very fast folding essentially takes the pattern from the unique low-$\rho$ state to high-$\rho$ with quenched geometry and then relaxes to the nearest minimum. Thus, as suggested by Figure~\ref{fig:RateDependence}(a), fast folding from the unique low-$\rho$ minimum picks out the glassy landscape configuration with closest geometric resemblance to the low-$\rho$ minimum. 

These considerations suggest an intriguing possibility - programming the bifurcation diagram using stiff creases can program different folding pathways that are followed at different rates of folding.

We tested this hypothesis on a $16$-vertex pattern with LP springs whose unique low-$\rho$ mode has significant geometric distortion relative to the adiabatic folding outcome; see Figure~\ref{fig:RateDependence}(b). We systematically folded this structure at increasing speeds relative to its hinge relaxation timescale. We find three completely distinct but reproducible folded structures in regimes of slow, intermediate and fast folding; see Figure~\ref{fig:RateDependence}(c). 

%Consequently, the speed of folding allows for dynamical control for targeting more than one folded configuration. 
%One can design a pattern and stiffness profile such that slow folding yields one folded structure, and fast folding leads to another. 

%Figure~\ref{fig:RateDependence}a compares these two scenarios side by side. We show how modifying the folding rate indeed leads to folding into different configuration for a large (3x3) origami pattern.

\section{Discussion}

%In this paper we argue for a general method of removing bifurcations in systems with one-dimensional soft modes. Introducing heterogeneous stiffness with relaxed states at the bifurcation point is shown to lift all but one undesired motion for both mechanical linkages and origami vertices. 

In this paper, we argued that meta-materials design should be conceptualized as targeted design of an entire dynamic pathway that avoids undesired behaviors, and not just target a desired final state. We showed how such pathways and their topological connectivity can be programmed by controlling the bifurcation diagram; we applied our method to remove the exponentially many misfolding motions intrinsic to self-folding origami.
 
%We related such pathways to bifurcation diagrams and showed simple ways to program such a bifurcation diagram 
%This idea of off-target design is effective at lifting all but one motion even if the branch point connects exponentially many motions, as in self-folding origami.

We showed that the bifurcation diagram can be modified by stiff joints, an inevitable feature of most experimental realizations of origami, linkage networks and other meta-materials. Thus, our proposal is conservative - it does not require specific directional information at hinges \cite{Peraza-Hernandez2014-gp}, temporal staging \cite{hawkes2010programmable} or using non-flat sheets \cite{Huffman1976-rr}.  Our general approach applies to any other heterogeneous bulk imperfection that couples to different folding modes unequally. 

%We discuss simple polynomial time algorithms, such as Linear Programming, for selecting the stiffness profiles.

%We further show how the convex landscape close to the bifurcation point is likely to connect continuously to the non-linear motion designed for the system. 

%This idea of heterogeneous stiffness and slow folding constitutes a useful heuristic for successfully folding self-folding origami, a problem believed to be NP-hard. 

%1. Mechanical meta-materials have targeted many interesting structures. 

%in order to achieve a desired goal. 
% Changed the
% Undesired pathways were made inaccessible from the unstrained state by eliminating them at saddle-node bifurcations. 

%2. Materials implications 

%3. Rate dependent outcomes .. different geometries, different mechanics.. 
A particularly intriguing direction suggested by our work is the ability to geometrically program different behaviors at different rates. These outcomes can have independently tuneable mechanical properties, such as energy absorption \cite{Paper1}. While such complex rate-dependent phenomena are actively studied in materials (e.g., cornstarch \cite{brown2010generality,lin2016tunable}), our approach suggests that rate-dependent behaviors can be dictated simply by the bifurcation geometry of the meta-material.

%For example, the different modes could have different , effectively creating a meta-material spring whose stiffness depends on  
%Our work suggests that different rates of actuation could evoke different structures 
% We leave

%4. 

\section*{Acknowledgements}

We thank Alfred Crosby, Heinrich Jaeger, Sidney Nagel, Jiwoong Park, and Thomas Witten for insightful discussions. We acknowledge NSF-MRSEC 1420709 for funding and the University of Chicago Research Computing Center for computing resources.

%unhandled text below%
%\bibliographystyle{unsrt}
%\bibliography{Paperpile_-_May_10_BibTeX_Export,NachiCitations}

\end{document}